\documentstyle[12pt]{article}
\newcommand{\stk}[1]{\stackrel{*}{\overline}}

\begin{document}
\begin{center}
\vfill
\large\bf{Supersymmetry in Three Dimensions}
\end{center}
\vfill
\begin{center}
D.G.C. McKeon$^\dagger$\\
Department of Applied Mathematics\\
University of Western Ontario\\
London\\
CANADA\\
N6A 5B7\vspace{.3cm}\\
T.N. Sherry$^{\dagger\dagger}$\\
Department of Mathematical Physics\\
National University of Ireland\\
Galway  IRELAND
\end{center}
\vfill
$^\dagger$email: DGMCKEO2@UWO.CA \\
$^{\dagger\dagger}$email: TOM.SHERRY@NUIGALWAY.IE
\eject
\section{Abstract}
The supersymmetry (SUSY) algebra and superspace in three
dimensions are examined in both Minkowski (3dM) and Euclidean (3dE)
space. Representations of the algebra are found and the implications of
requiring the norm of states in the Hilbert space is determined. Models are
described using superfield actions. The relationship of these models with supersymmetric
models in four dimensional Minkowski (4dM) space is described. The similarity
between $N = 2$ SUSY in 3dM and $N = 1$ SUSY in 3dE is noted.
\section{Introduction}
Supersymmetry has been widely discussed in 4dM [1-4]; it is anticipated
that it will eventually be shown to be a fundamental symmetry of nature.
The structure of the SUSY algebra is contingent of the nature of the
space in which it is defined. A discussion of the SUSY algebra in 4dE is
in [5-7]; 2 + 2 dimensions is analyzed in [8-11]. Scalar models with $N
= 1$ SUSY are introduced in [3] while some of their quantum properties
are worked out in [12-13]. Other SUSY algebras are considered in [14].

In this paper, we consider SUSY in 3dM and 3dE spaces.  In 3dM both $N =
1$ and $N = 2$ SUSY algebras are considered and we work out
representations of these algebras. A similar analysis is performed with
the simplest SUSY algebra in 3dE.  A superspace is introduced in
conjunction with each of these spaces and models worked out using
superfields defined in these superspaces. The relationship with SUSY in
4dM is considered.
\section{The SUSY Algebra}
In $2 + 1$ dimensions, we take the Dirac matrices to be
$$\gamma^0 = \left(\begin{array}{cc}
0 & -i\\
i & 0\end{array}
\right)\;\;\;\;\;
\gamma^1 = \left(\begin{array}{cc}
0 & i\\
i & 0\end{array}
\right)\;\;\;\;\;
\gamma^2 = \left(\begin{array}{cc}
i & 0\\
0 & -i\end{array}
\right)\eqno(1)$$
so that if [14]
$$A\gamma^\mu A^{-1} = \gamma^{\mu\dagger}\eqno(2a)$$
$$C\gamma^\mu C^{-1} = -\gamma^{\mu T}\eqno(2b)$$
we can make the identification
$$C = A = \gamma^0 .\eqno(3)$$
A spinor $\psi$ which transforms as
$$\psi \rightarrow U\psi\eqno(4a)$$
under a Lorentz transformation will also have
$$\overline{\psi} \rightarrow \overline{\psi} U^{-1}\eqno(4b)$$
$$\psi_C \rightarrow U\psi_C\eqno(4c)$$
if
$$\psi_C = C\overline{\psi}^T\eqno(5a)$$
and
$$\overline{\psi} = \psi^\dagger A.\eqno(5b)$$
It is possible to impose the Majorana condition
$$\psi_C = \psi = -\psi^*\eqno(6)$$
on a spinor in 3dM as $(\psi_C)_C = \psi$.

For $N = 1$ SUSY in 3dM we extend the usual Poincar\'{e} algebra by
introducing a Majorana spinorial generator $Q$ which satisfies that
anticommutation relation
$$\left\lbrace Q, \overline{Q} \right\rbrace = \gamma \cdot p .\eqno(7)$$
If we turn to the $N = 2$ algebra in 3dM, there are two Majorana spinorial
generators $Q_i$ ($i = 2, 2$) with
$$\left\lbrace Q_i , \overline{Q}_j \right\rbrace =
\gamma \cdot p\, \delta_{ij} + i \epsilon_{ij} Z\eqno(8)$$
where $\epsilon_{ij} = -\epsilon_{ji}$, $\epsilon_{12} = 1$ and $Z$ is a
central charge which commutes with all other generators.

When one works in 3dE, it is possible to identify the Dirac matrices
with Pauli spin matrices $\vec{\tau}$.
The solutions to eqs. (2a) and (2b) are now
$$A = 1\eqno(9)$$
$$C = \gamma^2\eqno(10)$$
so that now
$$\overline{\psi} = \psi^\dagger\eqno(11a)$$
$$\psi_C = C\overline{\psi}^T .\eqno(11b)$$
Since now $(\psi_C)_C = -\psi$ it is not possible to have a Majorana
spinor in 3dE. Spinors and Dirac, or equivalently, one may have a pair
of symplectic Majorana spinors $\psi$, and $\psi_2$ such that
$(\psi_1)_C = \psi_2$, $(\psi_2)_C = -\psi_1$. If we consider the
simplest possible supersymmetric extension of the Poincar\'{e} algebra
in 3dE we have Dirac spinorial generators $R$ such that
$$\left\lbrace R, (\overline{R}) \right\rbrace = \tau \cdot p +
Z\eqno(12)$$
where $Z$ is again a central charge.

Following the approach used in $3 + 1$ dimensions, we can decompose the
spinorial generators $Q$, $Q_i$ and $R$ in (7), (8) and (12) into
Fermionic creation and annihilation operators.  Beginning with $Q$, the
condition of eq. (6) shows the $Q$ is of the form
$$Q = i\left(\begin{array}{c}
q_1 \\
q_2\end{array}\right)\eqno(13)$$
where $q_1$, $q_2$ are real. If now
$$\Lambda = q_1 + iq_2\eqno(14)$$
then in the reference frame where $p^\mu = (M, 0, 0)$, (7) becomes
$$\left\lbrace \Lambda , \Lambda \right\rbrace > 0 \eqno(15a)$$
$$\left\lbrace \Lambda , \Lambda^\dagger \right\rbrace = 2M
,\eqno(15b)$$
showing the $\Lambda$ is a Fermionic annihilation operator. Next, upon
taking
$$Q_1 = i\left( \begin{array}{c}
\alpha_1\\
\beta_1 \end{array}\right)\;\;\;\;
Q_2 = i\left( \begin{array}{c}
\alpha_2\\
\beta_2 \end{array}\right)\eqno(16)$$
we can define
$$2\Lambda = \left(\alpha_1 - \beta_2 \right) + i
\left(\alpha_2 + \beta_1\right)\eqno(17a)$$
$$2 \Xi = \left(\alpha_1 + \beta_2 \right) + i
\left(\alpha_2 - \beta_1\right)\eqno(17b)$$
to that
$$\left\lbrace \Lambda , \Lambda^\dagger \right\rbrace = M -
Z\eqno(18a)$$
$$\left\lbrace \Xi , \Xi^\dagger \right\rbrace = M + Z\eqno(18b)$$
with all other anticommutators involving $\Lambda$, $\Xi$ being zero. In
order for the Hilbert space generated by these operators to be positive
definite, by (18) we must have [15-18]
$$M \geq Z.\eqno(19)$$

With the algebra of eq. (12), if
$$R = \left( \begin{array}{c}
r_1\\
r_2
\end{array}\right)\eqno(20)$$
then in the frame of reference where $\vec{p} = (0, 0, p)$, we see that
$$\left\lbrace r_1 , r_1^\dagger \right\rbrace = M + Z\eqno(21a)$$
$$\left\lbrace r_2 , r_2^\dagger \right\rbrace = -M + Z.\eqno(21b)$$
This shows that a positive definite Hilbert space for a supersymmetric
model in 3dE requires
$$M \leq Z.\eqno(22)$$
By comparing this result with eq. (19), we see that despite superficial
similarities, there are considerable differences between the algebra of
(8) and (12).  A similar difference occurs between the SUSY algebra in
4dM and 4dM [5-7].

Actual representations of the SUSY algebra of eq. (12) in 3dE are easily
found as in $3 + 1$ dimensions.  The operators $\vec{p}^{\;2}$, $\vec{J}^{\;2}$,
$\vec{P} \cdot \vec{J} /|\vec{P}|$, $Z$, $R$ and $R^\dagger$ are used to
classify the states. We begin with an initial state $|I>$ with
$$\vec{p}^{\;2} |I>\,=\,M^2|I>\eqno(23a)$$
$$\left(\vec{p} \cdot\vec{J}/|\vec{p}|\right) |I>\,=\,m|I>\eqno(23b)$$
$$\vec{J}^2 |I>\,=\,j(j+1)|I>\eqno(23c)$$
$$R |I>\,=\,0.\eqno(23d)$$
($\vec{J}$ is the angular momentum operator; consequently
$$\left[ J_i, J_j \right] i\epsilon_{ijk} J_k\eqno(24)$$
$$\left[ J_i, Q\right] = -\frac{1}{2} \tau_i Q.)\eqno(25)$$
Additional states in the representation are
$$r_i^\dagger |I> = |i>\eqno(26a)$$
$$r_1^\dagger r_2^\dagger |I> = |F>.\eqno(26b)$$
If we align $\vec{p}$ so that $\vec{p} = (0, 0, M)$, then it is easily
shown that
$$J_3 |1> \,=\, (m + \frac{1}{2})|1>\eqno(27a)$$
$$J_3 |2> \,=\, (m - \frac{1}{2})|2>\eqno(27b)$$
$$J_3 |F> \,=\, m|F>\eqno(27c)$$
and, as $\left[J^2, r_1^\dagger r_2^\dagger \right] = 0$,
$$J^2 |F> = j(j+1)|F>.\eqno(28)$$
The states $|1>$ and $|2>$ are superpositions of eigenstates of the
operator $\vec{J}^1$ corresponding to the eigenvalue $j$ shifted by
$\pm\frac{1}{2}$.

We now turn to a superspace formulation of these models.

\section{Superspace in Three Dimensions}
The simplest superspace we consider is the $N = 1$ SUSY algebra in 3dM
of eq. (7). Superspace consists of the usual Bosonic coordinates $x^\mu$
supplemented by a two component Majorana spinor $\theta$. The
supersymmetry algebra of (7) can be represented in this superspace by
the operators
$$Q = \frac{\partial}{\partial\theta} + \frac{1}{2} \overline{\theta}
\gamma \cdot p = C\overline{Q}^T\eqno(29a)$$
and
$$p^\mu = -i\partial^\mu .\eqno(29b)$$
(In [3, 12, 13] two component van der Waerden notation was used for the
spinors associated with this superspace.) The operator
$$D = \frac{\partial}{\partial\theta} - \frac{1}{2} \overline{\theta}
\gamma \cdot p\eqno(30)$$
anticommutes with $Q$.

We now can form a scalar superfield $\Phi(x,\theta)$; as $\theta$ has
two independent components an expansion of $\Phi$ in powers of $\Phi$
takes the form
$$\Phi(x, \theta) = A(x) + \overline{\lambda}(x)\theta +
F(x)\overline{\theta}\theta.\eqno(31)$$
(A term involving $\overline{\theta} \tau^a \theta = -\theta^T
C\tau^a\theta$ vanishes as $(C\tau^a)^T = C\tau^a$.) Generting
supersymmetric theories, both in terms of the superfield $\Phi$ and the
component fields $A$, $\lambda$ and $F$ is dealt with in detail in [3,
12, 13].

Now let us turn to $N = 1$ SUSY in 3dE. The algebra of eq. (12). We now
have two independent Fermionic spinors $\theta$ and $\theta^\dagger$,
each with two components, and hence the construction of a superspace
involves techniques similar to those employed in 4dM [19]. The algebra
of (12) can be realized by
$$R_i = \partial_i^\dagger - \frac{i}{2} \left(\tau^\mu \theta\right)_i
\partial^\mu - \frac{i}{2} \theta_i \partial_*\eqno(32a)$$
$$R_i^\dagger = \partial_i - \frac{i}{2} \left(\theta^\dagger \tau^\mu\right)_i
\partial^\mu - \frac{i}{2} \theta_i^\dagger \partial_*\eqno(32b)$$
$$p^\mu = -i\partial^\mu\eqno(32c)$$
$$Z = -i\partial_*\; \; .\eqno(32d)$$
(Notation and conventions used in 3dE appear in the appendix.)

Operators which commute with $R$ and $R^\dagger$ are
$$D_i = \partial_i^\dagger + \frac{i}{2} (\tau^\mu \theta)_i
\partial^\mu + \frac{i}{2} \theta_i\partial_*\eqno(33a)$$
and
$$D_i^\dagger = \partial_i + \frac{i}{2} (\theta^\dagger\tau^\mu)_i
\partial^\mu + \frac{i}{2} \theta_i^\dagger\partial_*\;\; .\eqno(33b)$$

If $\Phi = \Phi\left(x^\mu , \zeta, \theta_i, \theta_i^\dagger\right)$
is a scalar superfield then it forms an irreducible representation of
the algebra of (12) if it satisfies the ``chiral'' condition
$$D_i\Phi = 0 = D_i^\dagger \Phi^* .\eqno(34)$$
Noting that
$$D_i\theta_j = 0\eqno(35a)$$
$$D_i\left(x^\mu -\frac{i}{2} \theta^\dagger \tau^\mu\theta\right) = 0
\equiv D_i y^\mu\eqno(35b)$$
$$D_i\left(\zeta - \frac{i}{2} \theta^\dagger \theta \right) = 0 \equiv
D_iW\eqno(35c)$$
we see by (34) that
$$\Phi\left(x^\mu , \zeta , \theta_i , \theta_i^\dagger\right) =
\phi\left(y^\mu , W\right) + \lambda^\dagger \left(y^\mu, W\right)\theta
+ F\left(y^\mu , W\right)\theta_C^\dagger \theta\;\; .\eqno(36)$$

It is now possible to formulate a model in terms of chiral superfields as is done
in 4dM. A kinetic term is given by
$$S_k = \int d^3 x\, d\zeta d^2\theta d^2\theta^\dagger
\Phi^*\Phi\eqno(37a)$$
and a ``super potential'' by
$$S_p = \int d^3x d\zeta d^2\theta d^2\theta^\dagger \delta
|\theta^\dagger\left[m\Phi^2 + g_3 \Phi^3 + g_4\Phi^4\right] + {\rm{
H.C.}}\;\; .\eqno(37b)$$
Upon making the expansions
$$\Phi(y,w) = \phi(x, \zeta) - \phi_{,\mu}(x,\zeta)\left(\frac{i}{2}
\theta^\dagger \tau^\mu \theta\right) - \phi_*(x,\zeta)\left(\frac{i}{2}
\theta^\dagger \theta\right)\nonumber$$
$$+ \frac{1}{2!}
\phi_{,\mu\nu}(x,\zeta)\left(\frac{i}{2}
\theta^\dagger\tau^\mu\theta\right)\left(\frac{i}{2} \theta^\dagger\tau^\nu\theta\right)
\nonumber$$
$$+ \frac{1}{2!}
\phi_{,**}(x,\zeta)\left(\frac{i}{2}
\theta^\dagger\theta\right)^2 + \frac{1}{1!1!} \phi_{,*\mu}(x,\zeta)
\left(\frac{i}{2} \theta^\dagger\theta\right)
\left(\frac{i}{2} \theta^\dagger \tau^\mu\theta\right)\nonumber$$
$$+ \left(\lambda^\dagger(x,\zeta) - \lambda_{,\mu}^\dagger
(x,\zeta)\left(\frac{i}{2}\theta^\dagger\tau^\mu\theta\right) -
\lambda_*^\dagger (x,\zeta)\left(\frac{i}{2}
\theta^\dagger\theta\right)\right)\theta\nonumber$$
$$+ F(x,\zeta) \theta_C^\dagger \theta\eqno(38)$$
and using the formulae in the appendix for integrating over Grassmann
variables we find that (37) reduces to
$$S_k = \int d^3x d\zeta \left[-\frac{1}{16}\left(\nabla^2\phi^*\phi +
\phi^*\nabla^2\phi - 2\phi^*_{,\mu} \phi_{,\mu}\right)\right.\nonumber$$
$$-\frac{1}{16} \left(\phi_{**}^* \phi + \phi^*\phi_{**} -
2\phi_*^*\phi_*\right)\eqno(39a)$$
$$\left. -\frac{i}{8} \left(\lambda_{,\mu}^\dagger \tau^\mu \lambda +
\lambda^\dagger\lambda_* - \lambda^\dagger \tau^\mu\lambda_{,\mu} -
\lambda_*^\dagger \lambda\right) + F^2\right]\nonumber$$
and
$$S_p = \int d^3x d\zeta \left[ m\left(2\phi F + \frac{1}{2}
\lambda^\dagger \lambda_C\right) + g_3\left(3\phi^2 F + \frac{3}{2}
\phi\lambda^\dagger \lambda_C\right)\right.\eqno(39b)$$
$$\left. +g_4\left(4 \phi^3 F + 3\phi^2 \lambda^\dagger \lambda_C\right) +
({\rm{H.C.}})\right].\nonumber$$
It is interesting to note that the spinor $\lambda$ has Majorana, not
Dirac, mass and couplings.

To find the form of the SUSY transformations of the component fields, we
note that SUSY transformations in superspace are generated by the
unitary operator
$$U = \exp\left(\xi^\dagger R - R^\dagger\xi\right),\eqno(40)$$
and hence
$$\delta\Phi = \left[\xi^\dagger R - R^\dagger \xi ,
\Phi\right].\eqno(41)$$
From (32) and (36) we see that
$$\delta\phi = \lambda^\dagger\xi\eqno(42a)$$
$$\delta\lambda^\dagger = -i\xi^\dagger\left(\phi_{,\mu} \tau^\mu +
\phi_*\right) - 2F\xi_C^\dagger\eqno(42b)$$
$$\delta F = \frac{i}{2} \left(\lambda_{,\mu}^\dagger \tau^\mu 
\xi_C - \lambda_*^\dagger \xi_C\right).\eqno(42c)$$

It is also possible to formulate supersymmetric vector theories in 3dF.
We consider the case of a non-interacting $U(1)$ theory, again following
the techniques employed for the analogous model in 4dM.

We begin by noting that it is possible to form an irreducible representation
of the SUSY algebra of eq. (12) by imposing the reality condition
$$V = V^*.\eqno(43)$$
The most general form of this superfield (remembering eqs. (A.8) and
(A.9)) is
$$V = C(x,\zeta) + \left(\chi^\dagger\theta +\theta^\dagger\chi\right) +
\left(\theta^\dagger\tau^\mu\theta\right) V^\mu +
\left(\theta^\dagger\theta\right)E\nonumber$$
$$+ (M + iN)\theta_C^\dagger\theta + (M -
iN)\left(\theta^\dagger\theta_C\right) +
\left(\tilde{\Lambda}^\dagger\theta +
\theta^\dagger\tilde{\Lambda}\right)\left(\theta^\dagger\theta\right)\eqno(44)$$
$$ + \left(\theta^\dagger\theta\right)^2\tilde{D}.\nonumber$$
We now require invariance under the ``gauge transformation''
$$V \rightarrow V + \delta V = V + i\left(\Phi^* -
\Phi\right)\eqno(45)$$
as in 4dM, with $\Phi$ being a chiral superfield. In terms of component
fields, this means
$$\delta V^\mu = -\frac{1}{2} \left(\phi +
\phi^*\right)_{,\mu}\eqno(46a)$$
$$\delta E = -\frac{1}{2} \left(\phi + \phi^*\right)_*\eqno(46b)$$
$$\delta\tilde{D} = \frac{i}{8}\left(\nabla^2 -
\frac{\partial^2}{\partial\zeta^2}\right)(\phi^* - \phi)\eqno(46c)$$
$$\delta\tilde{\Lambda} = \frac{1}{2}\left(\tau \cdot \partial -
\frac{\partial}{\partial\zeta}\right)\lambda\eqno(46d)$$
$$\delta C = i(\phi^* - \phi)\eqno(46e)$$
$$\delta \chi = i\lambda\eqno(46f)$$
$$\delta(M + iN) = -iF \equiv \delta K.\eqno(46g)$$
We can now make the shifts
$$\tilde{\Lambda} =\Lambda - \frac{i}{2}\left(\tau \cdot \partial -
\partial_*\right)\chi\eqno(47a)$$
$$\tilde{D} = D + \frac{1}{8}\left( \nabla^2 -
\frac{\partial^2}{\partial\zeta^2}\right)C\eqno(47b)$$
so that (46c,d)
$$\delta\Lambda = 0 = \delta D .\eqno(48)$$
It is evident that $C$, $K$ and $M + iN$ can be eliminated by appropriate
choices of $\phi$, and $\lambda$ and $F$.  In this so-called ``Wess-Zumino'' 
gauge, $V$ reduces to
$$V_{WZ} = \left(\theta^\dagger \tau^\mu \theta \right)V^\mu +
\left(\theta^\dagger\theta\right)E + \left(\Lambda^\dagger\theta +
\theta^\dagger\Lambda\right)\left(\theta^\dagger\theta\right) +
D\left(\theta^\dagger\theta\right)^2\;\; .\eqno(49)$$

A field strength invariant under the transformation of eq. (45) is given
by the chiral spinor superfield
$$W_i = \left(D_C^\dagger D\right)D_{Ci}V.\eqno(50)$$
(Since $D^3 = 0$, we see that $D_j W_i = 0$.) An expansion analogous to
(36) leads to
$$W_i = X_i(y,w) + Y_{ij}(y,w)\theta_j + Z_i(y,w)\theta_C^\dagger\theta
.\eqno(51)$$
From (33) we see that in the Wess-Zumino gauge
$$X_i (x,\zeta) = W_i\left|_{\theta = \theta^\dagger = 0} =
2\Lambda_i\right.\eqno(52a)$$
$$Y_{ij}(x,\zeta) = D_j^\dagger W_i\left|_{\theta = \theta^\dagger = 0}
= 4D\delta_{ij} - 2\left(\vec{\tau} \cdot \vec{\nabla} \times \vec{V} +
i\vec{\tau} \cdot \vec{V}_* - i \vec{\tau} \cdot \vec{\nabla}
E\right)\right.\eqno(52b)$$
$$Z_i(x,\zeta) =
\frac{1}{4}\left(D^\dagger D_C\right)W_i\left|_{\theta=\theta^\dagger =
0} = i\left(\vec{\tau} \cdot \vec{\nabla} \Lambda_C -
\Lambda_{C*}\right)\right. .\eqno(52c)$$
Together, (51) and (52) show that the kinetic term
$$S_k = \int d^3x d\zeta d\theta
d\theta^\dagger\left[\delta\left(\theta^\dagger\right)
\left(W_i^\dagger\right)_C\left(W_i\right)\right] +
({\rm{H.C.}})\eqno(53)$$
reduces to
$$S_k = \int d^3x d\zeta \left[\Lambda_C^\dagger \tau \cdot \partial
\Lambda_C - \Lambda_C^\dagger\Lambda_{C*} + 4D^2\right.\eqno(54)$$
$$\left. - \left( \vec{\nabla} \times \vec{V} + i\vec{V}_* - i\vec{\nabla}
E\right)^2 + ({\rm{H.C.}})\right].\nonumber$$
This action is both SUSY and gauge invariant; coupling to chiral
``matter'' fields takes place in exactly the same manner as in 4dM. One
can also generalize to accommodate non-Abelian gauge symmetries as well.

Applying the analogue of (41) to the superfield $V$ in (44), we find the
SUSY transformation of the component fields of $V$ can be computed from
$$\delta V = \left[\xi^\dagger R - R^\dagger \xi , V\right].\eqno(55)$$
We find
$$\delta C = \xi^\dagger \chi + \chi^\dagger \xi\eqno(56a)$$
$$\delta E = \frac{1}{2} \left( \xi^\dagger\tilde{\Lambda} +
\tilde{\Lambda}^\dagger \xi \right) -
\frac{i}{4}\left[\chi_{,\mu}^\dagger \tau^\mu \xi + \xi_*^\dagger
\xi\right.\eqno(56b)$$
$$\left. - \xi^\dagger\tau^\mu \chi_{,\mu} - \xi^\dagger\chi_*\right]$$
$$\delta K = \frac{1}{2} \tilde{\Lambda}^\dagger \xi_C - \frac{i}{4}
\left(\xi^\dagger\tau^\mu \chi_{C,\mu} +
\xi^\dagger\chi_{C*}\right)\eqno(56c)$$
$$\delta V^\mu = -\frac{1}{2} \left(\xi^\dagger\tau^\mu\tilde{\Lambda} +
\tilde{\Lambda}^\dagger\tau^\mu\xi\right)\eqno(56d)$$
$$- \frac{i}{2}\left[\chi_{,\lambda}^\dagger \tau^\lambda \tau^\mu\xi -\xi^\dagger
\tau^\mu \tau^\lambda \chi_{,\lambda}\right.\nonumber$$
$$\left. + \chi_*^\dagger \tau^\mu \xi - \xi^\dagger\tau^\mu\chi_*\right]
\nonumber$$
$$\delta\tilde{\Lambda}^\dagger = 2\tilde{D}\xi^\dagger - \frac{i}{2}
\left[- \xi^\dagger \tau^\mu \tau^\lambda V^\lambda_{,\mu} -
\xi^\dagger \tau^\mu V_*^\mu \right.\eqno(56e)$$
$$+ \xi^\dagger \tau \cdot \partial K + \xi^\dagger E_* - 2
\xi_c^\dagger \tau\cdot \partial K\nonumber$$
$$\left. + 2 \xi^\dagger_c K_* + \xi^\dagger \tau \cdot \partial
E\right]\nonumber$$
$$\delta\tilde{D} = \frac{i}{4}\left[\xi^\dagger\tau^\mu
\tilde{\Lambda}_{,\mu} + \xi^\dagger \tilde{\Lambda}_*\right] +
({\rm{H.C.}})\eqno(56f)$$
$$\delta\chi = \left(\tau \cdot V\xi + E\xi + 2K^*\xi_C + \frac{i}{2}
\tau \cdot \partial C\xi + \frac{i}{2} C_*\xi\right).\eqno(56g)$$
A ``curl multiplet'' can be formed as in 4dM by
considering
$\Lambda$, $\Lambda^\dagger$, $D$, $V_{\mu,\nu} - V_{\mu ,\nu}$ and
$E_{,\mu} - V^\mu_{,*}$,
as by (56)
$$\delta D = +\frac{i}{4} \xi^\dagger(\tau \cdot \partial +
\partial_*)\Lambda\eqno(57a)$$
$$\delta\Lambda = 2D \xi + \frac{i}{2}\left[+ i\vec{\nabla} \times
\vec{V} \cdot \vec{\tau} + V^\mu_{,\mu} + E_{,\mu}\right]\tau^\mu
\xi\;\; .\eqno(57b)$$

It is evident that the Wess-Zumino gauge of eq. (49) is not respected by
the SUSY transformations of (56).  However, if one follows these SUSY
transformations with gauge transformations (46) with
$$\phi - \phi^* = 0\eqno(58a)$$
$$i\lambda + (\tau \cdot V + E)\xi = 0\eqno(58b)$$
$$-iF + \frac{1}{2} \Lambda^\dagger \xi_C = 0,\eqno(58c)$$
the Wess-Zumino conditions are seen to be restored. The combined effect
os SUSY and gauge transformations when one is in the Wess-Zumino gauge
is to effect the change
$$\delta E = -\frac{1}{2} \phi_* + \frac{1}{2}\left(\xi^\dagger \Lambda
+ \Lambda^\dagger\xi\right)\eqno(59a)$$
$$\delta D = +\frac{i}{4} \left( \xi^\dagger \tau \cdot \partial \Lambda
+ \xi^\dagger \Lambda_*\right) + ({\rm{H.C.}})\eqno(59b)$$
$$\delta\Lambda^\dagger = -\xi^\dagger\vec{\tau} \cdot
\left(\vec{\nabla} \times \vec{V}\right) - \frac{i}{2}
\xi^\dagger\tau\cdot\vec{\partial} E\eqno(59c)$$
$$+ 2 D\xi^\dagger + i\xi^\dagger\vec{\tau} \cdot \vec{V}_*\nonumber$$
$$\delta V^\mu = -\frac{1}{2} \phi_{,\mu} - \frac{1}{2}
\left(\xi^\dagger \tau^\mu \Lambda + \Lambda^\dagger\tau^\mu
\xi\right)\eqno(59d)$$
where by (58a) $\phi$ is a real field.

When one constructs a superspace for the $N = 2$ SUSY algebra associated
with 3dM (as given by eq. (8)), it is convenient to define a Dirac
spinor
$$\sqrt{2} S = \left(Q_1 + i Q_2\right)\eqno(60)$$
from the two Majorana spinors $Q_1$ and $Q_2$ so that (8) becomes
$$\left\lbrace S, \overline{S}\right\rbrace = \gamma \cdot p + Z
.\eqno(61)$$
As this is identical in form to (12), a superspace can be now
constructed in much the same manner as with $N = 1$ SUSY in 3dE. It is
not necessary to introduce ``harmonic superspace'' as is done with $N =
2$ SUSY in 4dM [20] or with $N = 1$ SUSY in 4dE [21].

It is now appropriate to relate SUSY in three dimensions with SUSY in
4dM.

\section{Dimensional Reduction From 4dM}

We consider the relationship between the $N = 1$ SUSY algebra in 3dM
$$\left\lbrace Q_\alpha , Q_\beta \right\rbrace = 0 = \left[ P_\mu ,
Q_\alpha \right]\eqno(62a)$$
$$\left\lbrace Q_\alpha , \overline{Q}_{\dot{\beta}} \right\rbrace = 2
\sigma^{\mu}_{\alpha{\dot{\beta}}} P_\mu \eqno(62b)$$
$$\left[M_{\mu\nu} , Q_\alpha \right] = 
-i\left(\sigma_{\mu\nu}\right)_\alpha^{\;\;\;\;\beta}Q_\beta\eqno(62c)$$
and the SUSY algebras of eqs. (8) (or (61)) and (12).
(We use the notation and conventions of (4).)  Beginning with the
ISO(3,1) algebra
$$\left[P_\mu , P_\nu \right] = 0\eqno(63a)$$
$$\left[ M_{\mu\nu} , P_\lambda \right] = i\left(\eta_{\nu\lambda} P_\mu
- \eta_{\mu\lambda} P_\nu\right)\eqno(63b)$$
$$\left[ M_{\mu\nu} , M_{\lambda\sigma}\right] = i\left(\eta_{\nu\lambda} 
M_{\mu\sigma} + \cdots\right)\eqno(63c)$$
($\eta_{\mu\nu} = (+, -, -, -)$)

We see that we can form a subalgebra $ISO(3) \times U(1)$ by simply
discarding the generators $M^{0i}$ in (62) and (63). If we now note that
$$\left(\sigma_{ij}\right)_\alpha^{\;\;\;\beta} = -\frac{i}{2}
\epsilon_{ijk} \tau^k\eqno(64)$$
and define
$$J^i = \frac{1}{2} \epsilon^{ijk} M_{jk}\eqno(65)$$
so that (62c) becomes
$$\left[J_i , Q_a\right] = -\frac{1}{2}\left(\tau_i Q\right)_a
.\eqno(66)$$
As
$$Q_a^* = \overline{\phi}_{\dot{a}}\eqno(67a)$$
$$Q^a = \epsilon^{ab} \phi_b\eqno(67b)$$
$$\overline{\phi}^{\dot{a}} = \epsilon^{\dot{a}\dot{b}}
\overline{Q}_{\dot{b}}\eqno(67c)$$
with $\epsilon^{ab} = \epsilon^{\dot{a}\dot{b}} = -iC$, we see that (62)
can be rewritten as
$$\left\lbrace Q, Q^\dagger\right\rbrace = \vec{\tau} \cdot \vec{p} +
Z\eqno(68)$$
upon making the identifications
$$Q_a \rightarrow \sqrt{2} Q\eqno(69a)$$
$$\overline{Q}_{\dot{a}} \rightarrow \sqrt{2} Q^\dagger\eqno(69b)$$
$$\overline{Q}^{\dot{a}} \rightarrow -i\sqrt{2} Q_C\eqno(69c)$$
$$Q^a \rightarrow i\sqrt{2} Q_C^\dagger\eqno(69d)$$
$$P_\mu \rightarrow (Z, \vec{p})\eqno(69e)$$
$$\sigma^\mu_{\alpha\dot{\beta}} \rightarrow (1,
\vec{\tau}).\eqno(69f)$$
Since (12) and (68) are identical, we see that the $N = 1$ SUSY algebra
in 3dE is that of the SUSY algebra associated with the $ISO(3) \times
U(1)$ subgroup of ISO(3,1).  It is now possible to use the mappings of
(69a-d) more generally; these can be used to establish the connection
between any spinor in 4dM and 3dE. This establishes a connection between
the model of eq. (39) and the Wess-Zumino model in 4dM and of the model
of eq. (54) and super QED. In the later case one must also identify the
scalar field $E$ with the temporal component $A_0$ of the vector field
in 4dM. The operator $\displaystyle{\frac{\partial}{\partial\zeta}}$ introduced
in (32d) is nothing but what is denoted by $\displaystyle{\frac{\partial}{\partial t}}$
in 4dM prior to eliminating the boost operators $M^{0i}$.

The $N = 2$ SUSY algebra in 3dM given by (8) and (61) can similarly be
obtained by a dimensional reduction from 4dM to 3dM. The algebra of
(62b) can be rewritten as
$$\left\lbrace Q, \phi^\dagger\tau_2\right\rbrace = p_0 \tau_2 +
p_1\left(i\tau_3\right) + p_2 + p_3\left(-i\tau_1\right).\eqno(70)$$
If now we make the identifications
$$p_2 = Z\eqno(71a)$$
$$\overline{Q} = Q^\dagger\tau_2\eqno(71b)$$
$$\left(\gamma^0, \gamma^1, \gamma^2\right) = \left(\tau_2, i\tau_3, -
i\tau_2\right)\eqno(71c)$$
and excise the rotation operators $M^{2i}$, then (61) and (70) can be
identified.

The dimensional reduction of $N = 1$ SUSY in 6dM to form $N = 2$ SUSY in
4dM and $N = 1$ SUSY in 4dE is considered in refs. [3] and [6,7]
respectively. In these reductions, the extra two dimensions are simply
discarded and do not become central charges as in (69e) and (71a).

Integration over $\zeta$ in (39) and (54) provides an extra ``fourth
dimension'' in these 3dE models; this clearly affects renormalizability.
In the $N = 2$ models in 4dM considered in [24] this degree of freedom
associated with central charge is compactified and its eigenvalues
discretized.

\section{Discussion}

We have considered a variety of SUSY algebras in three dimensions, both
in Euclidean and Minkowski space. All of these models admit a superspace
in which one can formulate models using both superfields and component
fields. A perplexing feature of the model of eq. (12) is the peculiar
nature of the bound of eq. (22); its significance is not at all clear.

A related problem that we are considering is the supersymmetric
extension of the Galilean group in $3 + 1$ dimensions. This Galielean
group can be generated by a Wigner-In\`{o}n\`{u} contraction of the
ISO(3,1) group [25]; presumably a similar contraction on the
supersymmetric extension of the ISO(3,1) group gives a supersymmetric
extension of the Galilean group.

\section{Acknowledgements}

We would like to thank NSERC for financial help.
D.G.C. McKeon would like to thank the National University of Ireland
Galway where this work was done. R. and D. MacKenzie had helpful advice.

\section{Appendix}
In $3 + 0$ dimensions we identify the Dirac matrices with the Pauli spin
matrices
$$\tau^1 = \left(\begin{array}{cc}
0 & 1\\
1 & 0\end{array}\right)\;\;\;
\tau^2 = \left(\begin{array}{cc}
0 & -i\\
i & 0\end{array}\right) = C\;\;\;
\tau^3 = \left(\begin{array}{cc}
1 & 0\\
0 & -1\end{array}\right) \eqno(A.1)$$
so that
$$\tau^\mu \tau^\nu = \delta^{\mu\nu} +
i\epsilon^{\mu\nu\lambda}\tau^\lambda \;\;.\eqno(A.2)$$
$$C\tau^\mu C^{-1} = -\tau^{\mu T}\; .\eqno(A.3)$$
Fierz identities can be proven by using the relations
$$\tau_{ij}^\mu \tau^\mu_{kl} = 2\delta_{il}\delta_{kj} -
\delta_{ij}\delta_{kl}\eqno(A.4)$$
$$\tau_{ij}^\mu \delta_{kl} + \tau_{kl}^\mu\delta_{ij} = 
 \tau^\mu_{il}\delta_{kj} + \tau_{kj}^\mu \delta_{il}\eqno(A.5)$$
$$\epsilon^{abc}\tau_{ij}^b \tau^c_{kl} = 
i\left(\tau_{il}^a \delta_{kj} -
\tau_{kj}^a \delta_{il}\right).\eqno(A.6)$$
From these relations one sees that if $\theta$ is a Dirac spinor, and if
$$\theta_c = C\theta^{\dagger T}\eqno(A.7)$$
then the following relations hold
$$\left(\theta^\dagger \tau^\mu
\lambda\right)\left(\theta^\dagger\theta\right) = -\left(\theta^\dagger
\tau^\mu\theta\right)\left(\theta^\dagger\lambda\right),\;\;
\left(\xi^\dagger\theta\right)\left(\theta^\dagger\theta_c\right) = -
2\left(\xi^\dagger\theta_c\right)\left(\theta^\dagger\theta\right)\eqno(A.8)$$
$$\left(\theta^\dagger\tau^\mu\theta\right)\left(\theta^\dagger\tau^\mu\theta\right)
= -\delta^{\mu\nu}\left(\theta^\dagger\theta\right)^2 = -
\frac{1}{2}\delta^{\mu\nu}\left(\theta^\dagger_c
\theta\right)\left(\theta^\dagger\theta_c\right)\eqno(A.9)$$
$$\theta^\dagger_k \theta^\dagger_{ell} = -\frac{1}{2} C_{kl}
\theta^\dagger\theta_c , \theta_k\theta_l = -\frac{1}{2}
C_{kl}\theta^\dagger_c\theta \eqno(A.10)$$
$$\left(\overline{\Lambda}\theta\right)\left(\overline{\xi}\theta\right)
= \frac{1}{2} \left(\overline{\Lambda}\xi_c\right)
\left(\overline{\theta}_c\theta\right),\;\;\;
\left(\overline{\theta}\Lambda\right) \left( \overline{\theta}\xi
\right)= \frac{1}{2}
\left(\overline{\xi}_c \Lambda\right)
\overline{\theta}\theta_c\eqno(A.11)$$
$$\left(\overline{\theta}\Lambda\right)\left(\overline{\xi}\theta\right) =
 - \frac{1}{2}\left(\overline{\theta}\tau^\mu\theta\right)\left(\overline{\xi}\tau^\mu
\Lambda\right) - \frac{1}{2} \overline{\theta}\theta \overline{\xi}\Lambda .\nonumber$$
For Grassmann integration, we employ
$$\int d^2\theta \theta_i\theta_j = -\frac{1}{2} C_{ij} = \int d^2\theta^\dagger
\theta^\dagger_i\theta^\dagger_j\eqno(A.12)$$
so that
$$\int d^2\theta \theta^\dagger_c\theta = 1 = \int d^2\theta^\dagger
\theta^\dagger\theta_c .\eqno(A.13)$$
The Bosonic operators $p^\mu$ and $Z$ are represented by
$$p^\mu = -i \frac{\partial}{\partial x^\mu} = -i\partial^\mu\eqno(A.14)$$
and
$$Z = -i \frac{\partial}{\partial\zeta} = -i\partial_*\eqno(A.15)$$
respectively.  We also use the notation
$$\frac{\partial}{\partial x^\mu} f(x^\mu , \zeta) = f_{,\mu}\eqno(A.16)$$
$$\frac{\partial}{\partial \zeta} f(x^\mu , \zeta) = f_{,*}\;\; .\eqno(A.17)$$
For derivatives with respect to Grassmann coordinates, we denote
$$\frac{\partial}{\partial\theta_i} = \partial_i\eqno(A.18)$$
and
$$\frac{\partial}{\partial\theta^\dagger_i} = \partial^\dagger_i .\eqno(A.19)$$
\end{document}